\documentclass{sigchi}

\toappear{\scriptsize Permission to make digital or hard copies of all or part of this work for personal or classroom use is granted without fee provided that copies are not made or distributed for profit or commercial advantage and that copies bear this notice and the full citation on the first page. Copyrights for components of this work owned by others than ACM must be honored. Abstracting with credit is permitted. To copy otherwise, or republish, to post on servers or to redistribute to lists, requires prior specific permission and/or a fee. Request permissions from permissions@acm.org. \\
{\emph{CHI '20, April 25--30, 2020, Honolulu, HI, USA.} } \\
Copyright is held by the owner/author(s). Publication rights licensed to ACM. \\
ACM ISBN 978-1-4503-6708-0/20/04\ ...\$15.00.\\
http://dx.doi.org/10.1145/3313831.3376548}

\usepackage{balance}       % to better equalize the last page
\usepackage{graphics}      % for EPS, load graphicx instead 
\usepackage[T1]{fontenc}   % for umlauts and other diaeresis
\usepackage{txfonts}
\usepackage{mathptmx}
\usepackage[pdflang={en-US},pdftex]{hyperref}
\usepackage[dvipsnames]{color}
\usepackage{booktabs}
\usepackage{textcomp}
\usepackage{pgfplots}
\usepackage{graphicx,siunitx}
\usepackage{subcaption,booktabs,dcolumn,adjustbox}
\usepackage{enumitem}
\usepackage[nocompress]{cite}
\usepackage{dblfloatfix}

\usepackage{microtype}        % Improved Tracking and Kerning
\usepackage{ccicons}          % Cite your images correctly!

\def\plaintitle{
Characterizing Twitter Users Who Engage in Adversarial Interactions against Political Candidates}

\def\plainauthor{Yiqing Hua, Mor Naaman, Thomas Ristenpart}

\def\plainkeywords{Online harassment, Twitter, user behavior, political candidates}

\newcommand\harassers{adversarial users}
\newcommand\threshold{\tau}

\makeatletter
\def\url@leostyle{%
  \@ifundefined{selectfont}{
    \def\UrlFont{\sf}
  }{
    \def\UrlFont{\small\bf\ttfamily}
  }}
\makeatother
\def\pprw{8.5in}
\def\pprh{11in}

\definecolor{linkColor}{RGB}{6,125,233}
\hypersetup{%
  pdftitle={\plaintitle},
  pdfauthor={\plainauthor},
  pdfkeywords={\plainkeywords},
  pdfdisplaydoctitle=true, % For Accessibility
  bookmarksnumbered,
  pdfstartview={FitH},
  colorlinks,
  citecolor=black,
  filecolor=black,
  linkcolor=black,
  urlcolor=linkColor,
  breaklinks=true,
  hypertexnames=false
}

\begin{document}

\title{\plaintitle}

\numberofauthors{1}
\author{
  \alignauthor{Yiqing Hua, Mor Naaman, Thomas Ristenpart\\
    \affaddr{Cornell Tech, Cornell University}\\
    \email{yiqing@cs.cornell.edu, mor@jacobs.cornell.edu, ristenpart@cornell.edu}}\\
}

\maketitle

\begin{abstract}
Social media provides a critical communication platform for political figures, 
but also makes them easy targets for harassment.
In this paper, we characterize users who adversarially interact with political figures on Twitter using mixed-method techniques.
The analysis is based on a dataset of 400~thousand users' 1.2~million replies to 756
candidates for the U.S. House of Representatives in the two months leading up to
the 2018 midterm elections. 
We show that among moderately active users, adversarial activity is associated with decreased centrality in the social graph and increased attention to candidates from the opposing party.
When compared to users who are similarly active,
highly adversarial users tend to engage in fewer
supportive interactions with their own party's candidates and express negativity in their user profiles. 
Our results can inform the design of
platform moderation mechanisms to support political figures countering online
harassment.

\end{abstract}

% ACM Classfication

\begin{CCSXML}
<ccs2012>
<concept>
<concept_id>10003120.10003121</concept_id>
<concept_desc>Human-centered computing~Human computer interaction (HCI)</concept_desc>
<concept_significance>500</concept_significance>
</concept>
<concept>
<concept_id>10003120.10003121.10003125.10011752</concept_id>
<concept_desc>Human-centered computing~Haptic devices</concept_desc>
<concept_significance>300</concept_significance>
</concept>
<concept>
<concept_id>10003120.10003121.10003122.10003334</concept_id>
<concept_desc>Human-centered computing~User studies</concept_desc>
<concept_significance>100</concept_significance>
</concept>
</ccs2012>
\end{CCSXML}

\ccsdesc[500]{Human-centered computing~User studies}
\ccsdesc[500]{Human-centered computing~Empirical studies in HCI}

\keywords{Online Harassment; Twitter; User Behavior; Political Candidates}

\printccsdesc

\section{Introduction}

Social media is a natural place for political figures to connect to the
public~\cite{grant2010digital,larsson2014course}, both by broadcasting and --- as
importantly --- listening
to their constituents and others~\cite{crawford2009following}.  At the
same time, social media services have been struggling with the prevalence of
online abuse and harassment~\cite{jack,suck,amnesty,pew2017,pew2012}.
Politicians in particular are common targets of negativity online, with
reports consistently showing that politicians face
overwhelming amounts of online harassment~\cite{nyt,ukreport}. This trend has discouraged some from
engaging in conversations online~\cite{theocharis2016bad}, caused others to quit
seeking public office~\cite{nyt}, and may have chilling effects on those who
would otherwise engage in democracy and public service.
It is therefore crucial to understand how to 
ensure a reasonable level of discourse on social media
for political figures and the broader public.

Making sense of incivility is complicated, with no clear definitions of
what types of content may cross acceptable boundaries in which contexts. 
A small but growing body of research uses
data-driven analyses to increase our understanding of incivility towards
politicians~\cite{gorrell2018twits,adversarial,tucker}, complementary to similar studies
of harassment in other
contexts~\cite{chatzakou2017mean,chatzakou2017measuring,elsherief2018peer,maity2018opinion}.
These works use various definitions of ``uncivil'' behavior.
For our purposes, we broadly refer to messages as adversarial if they are intended to
hurt, embarrass, or humiliate a targeted individual, although specific guidelines and accurate judgment
of such a definition is challenging, as we expand on below.

In this work, we use a combination of automated tools and qualitative coding
techniques to provide the richest characterization to date of Twitter users who
engage in adversarial interactions with political candidates. Fueling this
analysis is a dataset we collected\footnote{Available at: \url{https://figshare.com/articles/U_S_Midterm_Election_Twitter_Dataset_2018/11374062}}, capturing Twitter interactions with political candidates during the run-up
to the U.S.~midterm election in November 2018.
Our dataset consists of all
tweets by 1,110 accounts manually validated as used by 756 candidates for the
U.S.~House of Representatives. 
Additionally, it includes 1.2~million replies to the
candidates' tweets made by 0.4~million unique Twitter users, and these users'
Twitter profile information.  We focus our analysis on replies to candidates' tweets. 
Twitter replies are, by default, visible to other users interacting with the candidates' tweets.

With this data, we first identify users who post adversarial replies to any of the candidates by using Perspective API~\cite{perspectiveapi},
a general language tool for toxic content detection.
We show that while a
significant number of adversarial interactions are generated by users who only interact with candidates a few times,
over 35\% of the adversarial replies are created by just
10\% of the users who repeatedly post adversarial contents towards candidates.
We also explore the factors that are associated with 
higher levels of adversarial interactions with candidates, showing that the less a
user is embedded in the political discussions with candidates in general, the more they
tend to use adversarial language when replying to candidates.

We then focus on users who engage in significantly more adversarial
activity than others. We perform qualitative coding to better understand how
highly-adversarial users (those who posted more than 10 adversarial tweets to
candidates in our dataset) are different than other user groups in
terms of the content they post and the information they expose in their user
profile. We show that, 
the highly adversarial users post more personal attacks and generate more
off-topic replies, namely ones that are irrelevant to the candidate's tweet. 
In addition, these users' Twitter profiles are more likely to contain
partisan and adversarial attacks, potentially foreshadowing their negative
activity.  Finally, we evaluate how well the automated
tool we used for much of our analysis achieves good agreement with our human coders, validating Perspective API's effectiveness in identifying highly-adversarial actors. If
anything, the methods we used slightly under-estimate the amount of adversarial
content these users shared. 

Our work characterizes adversarial social media users in an important societal context, those who target political candidates. 
To this end, we perform context-specific analysis and show findings that both extend, and conflict with, those from previous works that also focus on adversarial users on Twitter but in different settings~\cite{chatzakou2017measuring,chatzakou2017mean,ribeiro2018characterizing,elsherief2018peer}.
Our work provides a better understanding of the activities and characteristic of these
``adversarial users'',
which can help inform design of social platforms, policies
against adversarial interactions, as well as mechanisms to allow for better moderation
and interactions with political figures in social media.
\section{Related Work}

A significant amount of previous work has focused on measuring  and understanding adversarial behaviors online, including studies on victims'
experiences~\cite{vitak2017identifying,matias2015reporting} and attackers'
motivations~\cite{cheng2015antisocial,cheng2017anyone,maity2018opinion}.
Other studies have developed machine learning tools that can be deployed at a large scale to support detecting personal attacks~\cite{wulczyn2017ex,djuric2015hate,nobata2016abusive}. 
We focus on three themes of related work that are pertinent to our work here:
(1) studies of online harassment on Twitter, (2) characterizing abusive Twitter users, 
and (3) studies of adversarial interactions against political figures.

\textbf{Online harassment on Twitter.}
Twitter has a particularly severe problem with the amount of online harassment on the platform~\cite{matias2015reporting,geiger2016bot}.
Victims who are targets of ongoing harassment reported that frequent encounters with harassing content disrupts their day-to-day lives~\cite{mahar2018squadbox}.
Yet, various institutions~\cite{matias2015reporting,amnesty} have reported that harassment reports from victims often are not addressed properly.

\textbf{Detecting and characterizing abusive users on Twitter.}
A number of recent projects attempt to characterize and automatically identify abusive users on Twitter~\cite{chatzakou2017measuring,chatzakou2017mean,elsherief2018peer,ribeiro2018characterizing}.
Some of the features considered by these projects include Twitter account
metadata~\cite{ribeiro2018characterizing,chatzakou2017measuring}, personality
traits~\cite{elsherief2018peer}, tweeting patterns~\cite{maity2018opinion}, and social network structures~\cite{ribeiro2018characterizing,elsherief2018peer,chatzakou2017measuring,chatzakou2017mean} associated with abusive behaviors. 
These studies used computational methods to provide quantitative insights on behavioral patterns of abusive Twitter users in different contexts~\cite{chatzakou2017measuring,chatzakou2017mean,maity2018opinion,elsherief2018peer,ribeiro2018characterizing}. 
Many of these works sample abusive or hateful speech by searching for a dictionary of explicitly offensive words~\cite{maity2018opinion,elsherief2018peer,ribeiro2018characterizing},
or mentions of known harassment campaigns~\cite{chatzakou2017measuring,chatzakou2017mean}.
Our work differs from these previous efforts in several ways.  
First, we focus
on an important societal context --- political candidates --- constructing our data by collecting all Twitter interactions with this specific set of users.
We leverage unique aspects of this political context (e.g., political leanings, political engagement on social media) to better analyze adversarial activities. 
Further, we complement our quantitative measurements with context-specific analysis based on qualitative coding to provide a deeper and more detailed understanding of user behaviors.

\textbf{Adversarial interactions against politicians.} 
Adversarial interactions against political figures, including politicians and candidates, have been studied long before the existence of social media (e.g.,~\cite{dietz1991threatening}). 
Recent surveys at politicians from multiple countries show that most of them have experienced
both online and offline harassment~\cite{james2016aggressive,adams2009harassment,pathe2014harassment,every2015harassment}. 
The rise of social media provides political figures with an important channel to connect with citizens~\cite{parmelee2011politics,crawford2009following,agarwal2019tweeting},
as well as a platform where they can be easily addressed and targeted by online harassment~\cite{every2015harassment,gorrell2018twits,nyt,ukreport,adversarial}. 
It has been reported that social-media-based abuse discourages politicians from engaging in online conversations~\cite{theocharis2016bad}. At the extreme, some reported quitting seeking public office due to the amount of harassment on social media~\cite{nyt}.  
Naturally, many political figures have professional teams running their social media accounts, insulating them to some extent from harassment. 
However, this does not obviate the ill effects of adversarial content,
for example such content might have negative impact on other people who engage with the candidate's tweets~\cite{cheng2017anyone}. 

Closest to our work is research from Gorrell et
al.~\cite{gorrell2018twits,gorrell2018online}, examining Twitter harassment
towards United Kingdom parliament members with emphasis on
understanding what attributes of politicians -- such as party and gender -- and what
political topics attract abusive replies. 
Our own work used the same dataset
we analyze here to study how this set of U.S. House candidates experienced
harassment on Twitter in that time
period~\cite{adversarial}.
Both works focus on measuring trends in the adversarial interactions experienced by the
candidates,
such as the different forms that
adversarial interactions may take~\cite{adversarial},
or topics that adversarial interactions focus on~\cite{gorrell2018twits,gorrell2018online}.
In contrast, our focus here is on
characterizing the \textit{users} that are making adversarial posts.
\section{Data Collection and Analytic Approach}\label{sec:data}

We use a data-driven approach to gain a better understanding of the users who
post adversarial replies to political candidates online.  
We define adversarial interactions as 
messages posted by users which are intended to hurt, embarrass or humiliate a target user,
following a notion that has been largely accepted in the community~\cite{maity2018opinion,corcoran2015cyberbullying,dinakar2012common,dinakar2011modeling,singh2017they,grigg2010cyber,adversarial}.
Of course, determining whether a message satisfies this definition can be highly
subjective, and we will use a combination of existing machine learning techniques, and, in
a following section, human annotations, to assess whether messages are adversarial.

\textbf{Our dataset.}
We built on a set of $1,110$ Twitter accounts used by 786 candidates, 431 Democrats~(D) and 355
Republicans~(R) running in 2018 for the U.S. House of
Representatives developed in our earlier work~\cite{adversarial}.
Our study is focused on 1.2~million Twitter replies to
the candidates' posts, made by 0.4~million users between September 17th and
November 6th, 2018 (the day of the U.S. midterm elections). We collected the data
using the Twitter Streaming API, including all tweets posted by, replying to, or
retweeting any of the candidate accounts.  On Twitter, replies to public
accounts directly engage with the target account and are visible by default to the public
who are viewing the original posts.  Thus, replies to candidates directly
contribute to the candidate's own ``public sphere''~\cite{twitterReply}. We therefore focus
on replies to candidate posts in this work.
We do not include 
Twitter ``mentions'' of these accounts and tweets that include a link to tweets by these accounts (often called ``quote tweets''),
which may be visible to others on Twitter but are
not immediately available to those interacting with a candidate's account.
We alternatively refer to tweets in our dataset as ``replies'' or simply ``tweets''---but note that all of these are direct replies to tweets by the candidates.

We augment the dataset of replies with the Twitter
following network for the users in the dataset. On Twitter, analyzing who users
follow has proven to be informative for inferring user interests and political
preferences~\cite{romero2013interplay,barbera2015birds}.  We therefore retrieve
the 5,000 most-recently followed accounts by each user in our dataset (i.e., 
their ``Twitter friends'') using the Twitter API~\cite{twitterapi}.  Due to API rate limits, this data collection lasted
over a longer time period, from September 2018 until March 2019.  This network data is
limited in at least three ways. First, by the time of data collection, some users
have left Twitter or set their profiles to be private, preventing us from
retrieving their friends list.  However, since our data collection prioritized
the more active users in our dataset, we do not think these omissions are critical.
Second, users might follow new accounts between the time they made a reply to a candidate and their data being collected by us,
resulting in some inaccuracies.
Last, we were limited to retrieve only the 5,000 most recent followers from any user.
As a result, we obtained the complete friends list for $92\%$ of the users.
For an additional $4\%$ of users who
followed more than 5,000 accounts, we obtained the partial friends list.
The remaining $4\%$ of accounts
were deleted or
suspended at the time of our network data collection.

Note that it is possible that organized \textit{information operations} took place on Twitter during our data collection period,
whether using bots powered by automated algorithms~\cite{davis2016botornot,ferrara2016rise}
or through state-operated campaign~\cite{im2019still, arif2018acting}. 
We compared our data with a list of state controlled accounts with over 5,000 followers published by Twitter~\cite{twitter2018midterm}~(as the information of this set of accounts was not anonymized):~none of them showed up in our dataset.
Further, manual review of highly-active and highly-adversarial users as discussed later
indicates that these users exhibit sophisticated behaviors that were not likely generated by automated algorithms.
Finally, most of our analyses would remain relevant 
even if the accounts are automated or are part of organized campaigns since the resultant adversarial activities 
are nevertheless perceived by, and have impact on others, e.g. the candidates and the observers.

\textbf{Automated labeling of adversarial interactions.}
The scale of this dataset suggests the need for a computational approach to
analyze adversarial interactions. We use machine learning tools for flagging
toxic content in the first part of this work, and as a mechanism to identify content to analyze qualitatively in the second part of this work.
Specifically, we use Perspective API~\cite{perspectiveapi} to assign toxicity scores to tweets. 
The API has been used before in different contexts, including Twitter~\cite{adversarial,elsherief2018peer}. 
The scores generated by Perspective API indicate whether a general language model, trained with 
Wikipedia discussions and multiple other online conversation sources~\cite{wulczyn2017ex,perspectiveapi}, believes the utterance to be ``discouraging participation in conversations''. 
The API outputs a ``toxicity'' score in the range $[0,1]$, with~1 being highly toxic and~0 being non-toxic. 
We define a tweet as \emph{$\threshold$-adversarial} if its content has a toxicity score more than $\threshold$.
Note that although similar, toxicity is in fact a different concept than being adversarial~(see the definition we use at the beginning of this section),
as one can be toxic without being adversarial (e.g., using foul language as expression), and adversarial without being toxic (e.g., combative speech).
Here, we use toxicity in replies as an approximation of adversarial actions. 
In our earlier work, we validated the precision of this technique 
for \emph{$0.7$-adversarial} tweets,
showing this method offers high precision~\cite{adversarial}.

Nevertheless, Perspective API as well as other existing automated
approaches that use general language models to flag adversarial
content~\cite{djuric2015hate,nobata2016abusive}
have inherent limitations, in particular having potentially low recall for some forms of
adversarial interactions. 
We mitigate the limitations inherent in using a general language model in two key ways.
First, 
we provide analysis showing that the trends in our results are robust to the 
toxicity thresholds we use in our computational analysis. 
Second, we report on a large-scale qualitative coding of tweets 
that helps expose whether the method is prone to specific biases in the sources of abuse it detects.

\section{Volume of adversarial activity}\label{sec:volumne}

\begin{figure*}[ht]
    \includegraphics[width=\textwidth]{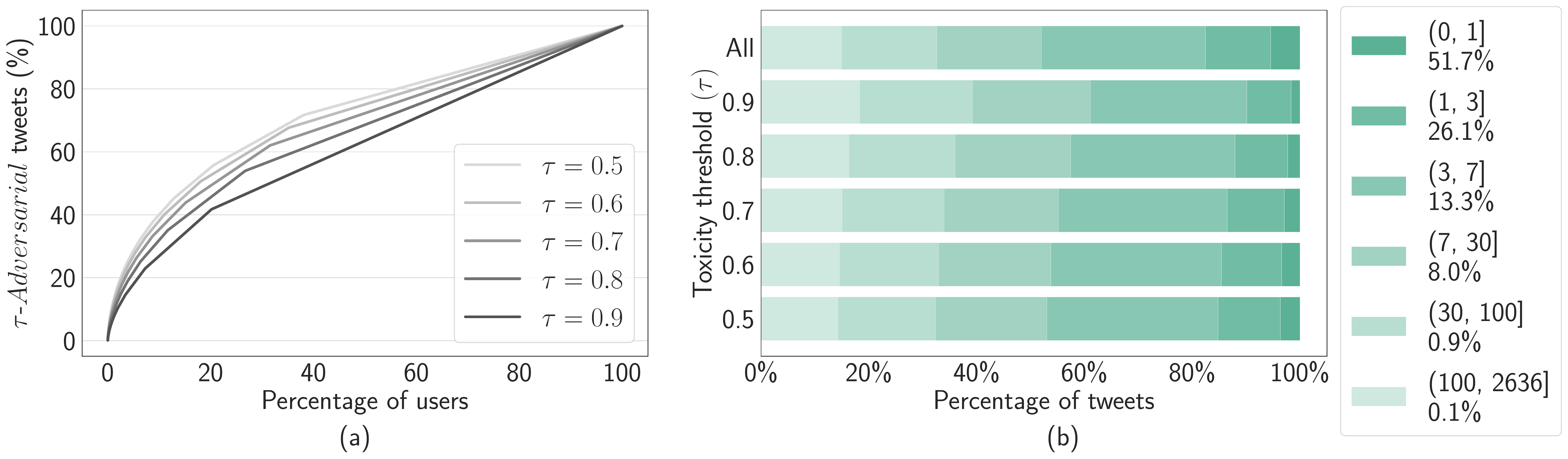}
\vspace{-.7cm}
\caption{
\label{fig:fig1}
\textbf{(a)} CDF of $\tau$-adversarial tweets contributed by users.
\textbf{(b)} Breakdown of replies by users with different \textit{overall} activity~(top row), with the fraction of adversarial tweets made by each user group, for various
$\tau$ (following rows).
The legend specifies user overall activity level for each group, with the group size in percentage.}
\end{figure*}
\begin{table}[!t]
    \small
   \centering
    \begin{tabular}[!t]{l|rrrrr}
      \toprule
      $\threshold$ &  0.5 & 0.6 & 0.7 & 0.8 & 0.9\\
      \midrule
      $\threshold$-adversarial tweets ($\%$) & 28\% & 21\% & 14\% & 9\% & 4\% \\
      Average per user & 2.2 & 2.0 & 1.8 & 1.6 & 1.3\\ 
      \bottomrule
   \end{tabular}
   \caption{\label{tab:threshold} Fraction of $\threshold$-adversarial
   tweets in our dataset, for varying thresholds $\threshold$, and the
   average number of $\threshold$-adversarial tweets made by a user that made at least one such tweet.
   }
\end{table}

To gain an initial understanding of the use of toxic language by
different users, we analyze the distribution of $\threshold$-adversarial
activities per user observed in our dataset. The analysis is geared towards exposing
the levels of activity of \harassers, and how those levels relate to the general activity
levels of users interacting with candidates. 
A secondary goal of the analysis
is to understand the sensitivity of these trends to the
values used for the threshold~$\tau$.
In general, changing the threshold results in a significant change to the number of tweets classified as toxic:  Table~\ref{tab:threshold} provides some indication on how raising the
threshold~$\threshold$ decreases the number of tweets considered
as toxic.  

Regardless of the specific $\threshold$,
a relatively small number of users generate a large fraction of adversarial
tweets, in line with prior studies on online communities~\cite{nonnecke2000lurker,preece2004top}.
Figure~\ref{fig:fig1}a shows the cumulative fraction of
$\threshold$-adversarial tweets (Y-axis) contributed by an increasing fraction
of the users (X-axis), for various values of $\threshold$.
Users are sorted in descending order of the number of $\threshold$-adversarial
tweets they made.
For example, $49\%$ of the $0.7$-adversarial tweets
were posted by $20\%$ of the users and $35\%$ of the $0.7$-adversarial tweets were
generated by just $10\%$ of the users. 
Note that the total number of tweets is different for
each curve, as indicated by Table~\ref{tab:threshold}.

Users who had fewer overall interactions with the candidates contribute more adversarial replies than
their overall contribution predicts.
Figure~\ref{fig:fig1}b breaks down the proportion of adversarial activity with different $\threshold$ (separated by rows),
contributed by users with different levels of \textit{overall} activity in our dataset (shades of green). 
The lightest shade (left) are users who are at or below the median (51.7\% of the users in our dataset that posted a total of 1 tweet reply to candidates). The following darker shades are the 75th percentile (2 or 3 tweets), $90\%$ (4--7 tweets), $99\%$ (8--30 tweets) and $99.9\%$ (31--100 tweets).
The most active $0.1\%$ (over 100 tweets) is in darkest shade of green.
The top row of the chart shows the breakdown of all tweets by these groups,
while the remaining rows show a breakdown of adversarial tweets by activity level groups
for different levels of $\threshold$-adversarial content.
For example, the figure shows that $51.7\%$ of users who replied to candidates only once (leftmost, lightest shade) contributed $15\%$ of all replies (top row), but over $18\%$ of the 0.9-adversarial replies (second from top).
Grouped together, the top $10\%$ most active users (the three rightmost, darkest groups) contributed $48\%$ of all replies (top bar) but only $39\%$ of the 0.9-adversarial replies (second from top bar).
In summary, while ``heavy'' adversarial users still carry most of the weight, Figure~\ref{fig:fig1}b shows that ``casual'' offenders reply more with adversarial content than their overall activity predicts.
\section{Characterization of Adversarial Users}
\label{sec:chracterizing}

We are interested in the tendency of users to share more adversarial content than expected given their basic user characteristics, engagement with political topics and adversarial behavior of Twitter friends.
To this end, we perform an analysis of a set
of moderately active users --- those who posted between 4 to~30 replies to candidates.  
The $[4,30]$ user group comprises only $21\%$
of the user base (N=74,533), but authored $50\%$ of all tweets and $52\%$ of the
adversarial tweets in our data.
As the distribution of user activity is highly skewed, this group excludes the under-~(users who produced fewer than 4 tweets) and over-producers~(users who produced more than 30 tweets) of content. 
The over-producers were removed because their behaviors are likely to be materially different than less active users.  
The under-producers do not have enough content in our dataset to provide meaningful analyses. 
We performed a similar analysis that included the under-producers in addition to the $[4, 30]$ user group,
and the regression results reflected a similar trend.

Specifically, we focus on three categories of features in our analysis:
the basic characteristics of the user as reflected in their Twitter account metadata; 
the specific characteristics of the user's Twitter ``political activity''~(in our data); 
and features that capture adversarial behaviors by a user's friends on Twitter.
We fit these features into a linear regression to the number of
$\threshold$-adversarial tweets by the user as calculated using Perspective API.
For simplicity, we use $\threshold=0.7$ and refer to
$0.7$-adversarial tweets as ``adversarial tweets'' from now on. We set this threshold 
based on prior work~\cite{adversarial} and validation of precision associated with the 0.7 threshold in the later discussion via manual labeling.
Also, as shown above, the analysis trends are robust to
small changes in $\threshold$.

\subsection{Regression Features}

\textbf{Basic user characteristics.}
We start with a set of basic attributes about users inspired by previous studies: their number of followers,
the age of their account,
whether their account is verified or not, and their level
of activity.
Prior works in other contexts have
shown that the number of followers is correlated with abusive behavior.
Interestingly, the effect is context dependent, with some studies indicating
that more followers a user has is positively~\cite{chatzakou2017measuring} correlated with adversarial
behavior while other studies show the opposite
trend~\cite{ribeiro2018characterizing,chatzakou2017mean}. 
Those prior works also use the age of an account (i.e., how long has this user been registered on Twitter), and, again, 
show differing correlations with adversarial
behavior~\cite{ribeiro2018characterizing,chatzakou2017measuring}. 
Previous studies on online forums show that anonymity can be used as a criteria
to distinguish attackers from defenders of aggressive
behavior~\cite{moore2012anonymity,ybarra2004youth}.
We do not have a robust
measure of anonymity of Twitter accounts, as even those who seem to be using
their real names could be using others' identities. Instead, we use the Twitter
``verified''~\cite{twitterVerified} status as a (limited) replacement. This feature is offered by Twitter to users who are public figures, and marks the fact that the user identity had been verified by Twitter.
Finally, we include the number of replies posted by user to candidates in our dataset as a control, 
to account for the heightened user activity level that may
give them more opportunity to behave in an uncivil manner.

\textbf{Engagement with political topics.}
A second set of features we explore captures various aspects of a user's political activity on Twitter, including interactions with candidates, as well as 
the candidate accounts followed by the user and the amount of retweeting candidates' posts.
We use the number of candidates followed and number of candidate tweets retweeted 
by a user as approximate measures of a user's interest in politics.
Using methods described in~\cite{adversarial}, we label each user as either pro-Democrat or pro-Republican based on their retweeting of candidates,
except for 1\% of the moderately active users for whom we do not have enough retweet information to classify partisanship, who are excluded from the analyses in this section.
Opinion conflicts are indicative features of incivility~\cite{maity2018opinion}. 
To measure the potential frequency of opinion conflicts,
we include both the number of candidates of the user's opposing 
party that the user has replied to, and the percentage of the user's replies
that were directed
towards candidates from the opposite party.

Another measure of the user's level of political engagement is in their
self-description. For example, some users include partisan hashtags such as
\#MAGA in their Twitter profile.  Using a list of partisan
hashtags compiled from~\cite{adversarial}, we identify users who express their
political preferences in their user profiles.  This feature is used in the
following analysis as a signal of a user's partisanship.
A final measure we use for political engagement is the
centrality of the users in the discussions involving the candidates. To
compute this measure, we calculate the page rank score~\cite{page1999pagerank} of each user in our
dataset in the network of the following relationship of all the users who reply
to political candidates (i.e., all users in our full dataset).
In other words, the network is the partial graph of the Twitter following network where
the set of nodes is limited to the users in our data.

We have also considered other factors, including absolute number of replies to
candidates from the opposing party, number of followers supporting
the same party, and number of candidates followed by the user from the party
they support --- but ultimately excluded them from the analysis due to
multicollinearity~\cite{o2007caution} with other variables we are already considering.

\textbf{Adversarial behavior of Twitter friends.}
It is known that a user's adversarial activity level may be associated with the activity of their Twitter friends~(i.e., the accounts a user is following on Twitter)~\cite{mcpherson2001birds,al2012homophily}.
We consider two attributes to measure this effect: the total number of adversarial tweets posted by a user's Twitter friends, and 
the percentage of one's friends that posted at least $x$ adversarial tweets.
We use $x=3$ in the following analysis, but verified that the results of $x=1,2,4,5$ are similar.

\subsection{Results}
\label{sec:regression_res}
\setlength{\tabcolsep}{4pt}
\begin{table}[!t]
    \small
    \begin{tabular}[t]{llllrr}
      \toprule
      & Model 1 & Model 2 & Model 3 & Mean & Std\\
      \midrule
      
      (Intercept) & 1.237\tiny{***} & 1.237\tiny{***} & 1.237\tiny{**} & &\\
                  & (0.006)  & (0.006)  & (0.006) & &\\\\
       
      Number of replies
      & 1.49*** & 1.314*** & 1.307*** & 8.14 & 5.21\\
      \includegraphics[width=.1\linewidth]{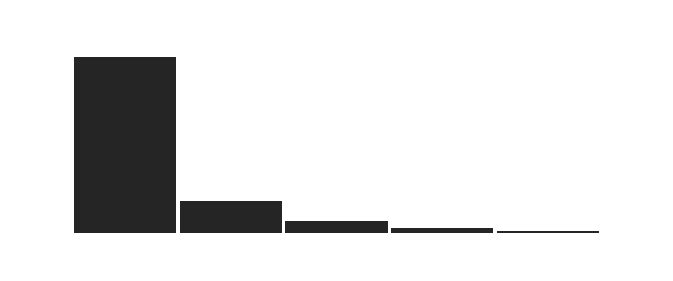}& (0.012) & (0.013) & (0.013) &&\\\\
      
      Account age (days)
      & -0.04** & -0.052*** & -0.004 &
      1818 & 1201\\
      \includegraphics[width=.1\linewidth]{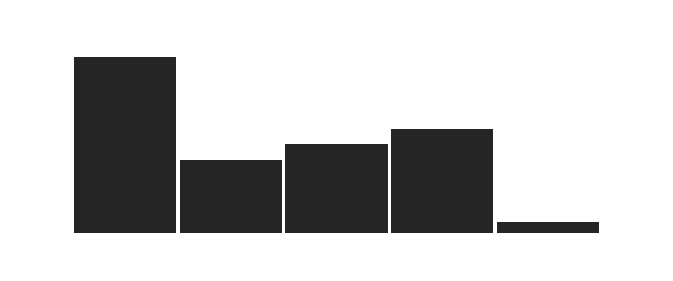} & (0.012) & (0.012) & (0.012) && \\\\
      
      \textit{Number of followers}
      & -0.052*** & -0.036* & -0.138*** & 
      2.36 & 0.91 \\
      \includegraphics[width=.1\linewidth]{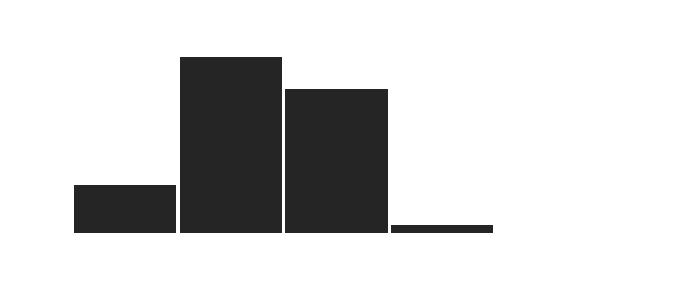} & (0.015) & (0.016) & (0.022) & & \\\\
      
      Verified on Twitter & -0.4** & -0.006 & -0.218 & & \\
      ($99.8\%$ not verified) & (0.132) & (0.012) & (0.136) & & \\
      \end{tabular}
      
     \begin{tabular}[t]{l@{\hskip 0.1in}llrr}     
      \midrule
      \textit{Number of retweets}
      & -0.032* & -0.044** & 0.71 & 0.67\\
      \includegraphics[width=.1\linewidth]{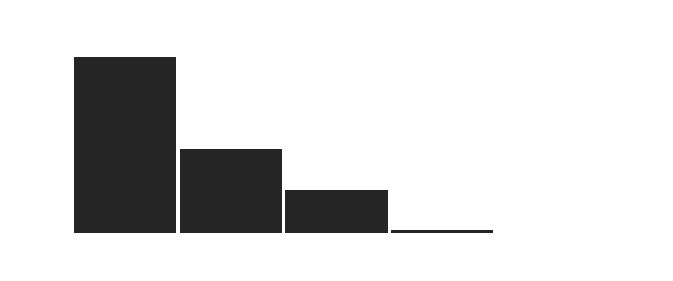}
      & (0.013) & (0.013) & & \\\\
      
      Candidates followed
       & -0.048*** & -0.051*** & 11.4 & 19.6\\
      \includegraphics[width=.1\linewidth]{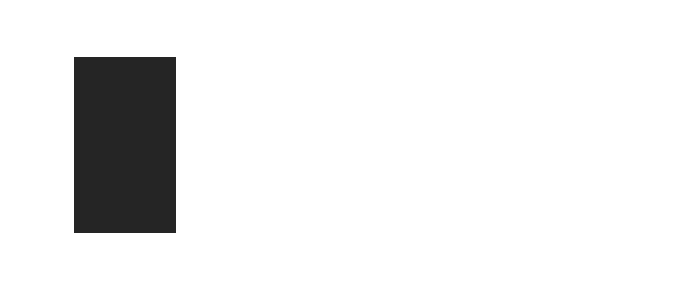}
       & (0.012) & (0.013) & &\\\\
       
      \textit{Page rank}
      & -0.068*** & -0.118*** & 0.09 & 0.08\\
      \includegraphics[width=.1\linewidth]{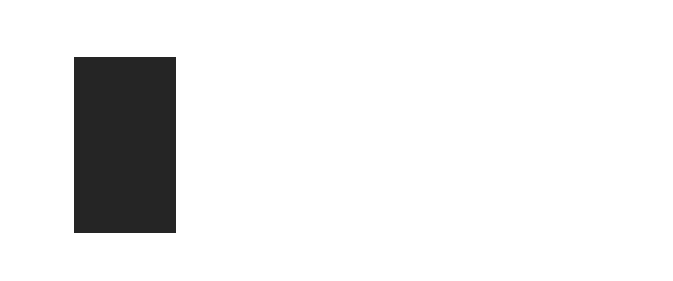}
      & (0.015) & (0.015) & & \\\\
      
      Partisan hashtag in profile (bool)
      & 0.085*** & 0.051** & & \\
      ($80\%$ don't have partisan hashtag)& (0.013) & (0.017) & &\\\\
      
      \% replies to opponent candidates
      & 0.429*** & 0.426*** & 53\% & 38\% \\
      \includegraphics[width=.1\linewidth]{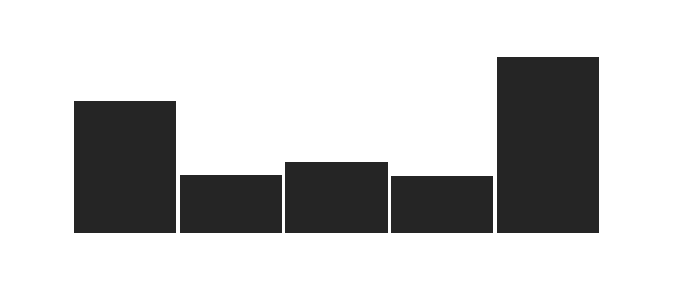}
      & (0.015) & (0.015) & & \\\\
      
     Opponent candidates replied to
      & 0.43*** & 0.433*** &
      2.14 & 1.93 \\     
     \includegraphics[width=.1\linewidth]{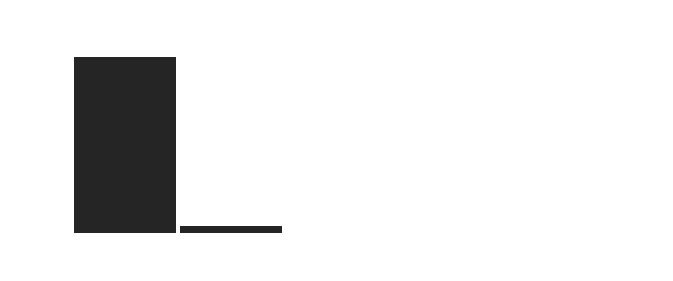}
      & (0.016) & (0.016) & & \\
     \end{tabular}
      
     \begin{tabular}[t]{l@{\hskip 0.4in}lrr}     
      \midrule
      Adversarial tweets posted by friends
      & 0.076** & 1.49 & 0.99\\
      \includegraphics[width=.1\linewidth]{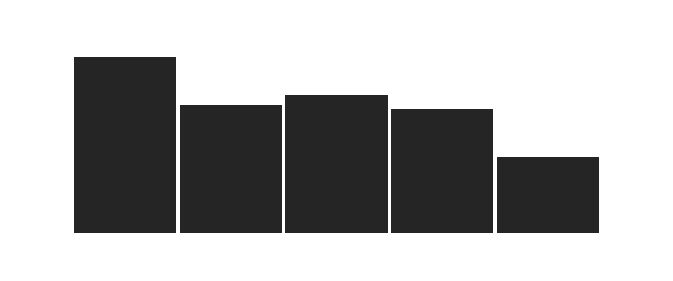}
      & (0.025) &&\\\\
      
      \% friends posted $\geq 3$ adversarial tweets
      & 0.204*** & 1.1\% & 1.2\% \\
      \includegraphics[width=.1\linewidth]{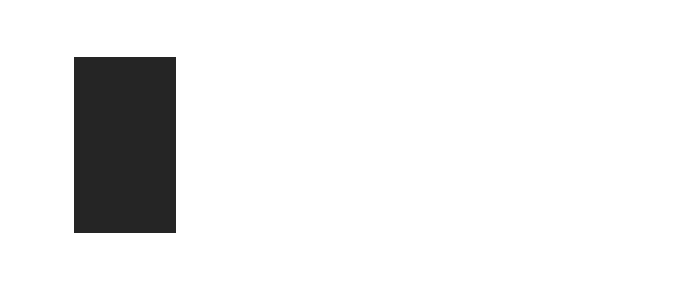}
      & (0.017) &&\\
      \end{tabular}
      
      \begin{tabular}[t]{l@{\hskip 1in}l@{\hskip 0.3in}l@{\hskip .3in}l@{\hskip .9in}}
      \midrule
      $R^2$ & 0.18 & 0.227 & 0.231\\
      \bottomrule
      \multicolumn{4}{l}{Significance codes: *** $p < 0.001$, ** $p < 0.01$, * $p < 0.05$}
      \end{tabular}
      \caption{\label{tab:regression} Results of the regression analysis predicting the number
      of adversarial tweets posted by a user~(number of users is $N=74,533$).
      Histograms represent the distribution of the variable.
      Log transformed variables are listed in italic,
      with mean, standard deviation and distribution in log scale.}
      \vspace{-.5cm}
\end{table} 

We analyze the power of the attributes listed above in predicting the dependent variable
--- the number of 0.7-adversarial tweets made by the user --- using three
increasingly complete linear regression models that incrementally add each set of features
described above. \emph{Model~1} uses the basic user Twitter
features (the first group described above),
including controlling for the user's overall
activity level. \emph{Model~2} combines these features with
the features capturing the user's political activity in our dataset.  
Finally, \emph{Model~3} adds the features capturing
adversarial behaviors of the user's Twitter friends.

The histograms in Table~\ref{tab:regression} show the data distribution of each variable used in the regression models,
along with its average and standard deviation (rightmost two columns).
As the table histograms show, not all variables are normally distributed. To account for this, we log-transform 
the number of followers, number of retweets, page rank score, and amount of adversarial activities of Twitter friends (the italic features in Table~\ref{tab:regression}).
Furthermore, all continuous
variables were standardized by centering and dividing by two standard deviations
following recommended practice~\cite{gelman2008scaling}.  
The binary variables were centered in order to have the same scale for comparison.
Hence the magnitude of the standardized coefficients reflects the significance of the variables.
Table~\ref{tab:regression} also shows the results of the regression analysis, 
including the standardized beta coefficients ($b$) and their standard errors
($SE$) for all three models. 
The $p$ values are computed using two-tailed t-tests.
All variance inflation factors~(VIFs) are less than three, indicating that
multicollinearity is not an issue in our independent variables.  Overall, the
three models 
explain $R^2 = 0.18$, $R^2 = 0.227$, and $R^2 = 0.231$ of the variance of the dependent
variable (bottom row of the table).

Among the features included in \emph{Model~1}, the number of replies has the
strongest correlation with adversarial replies ($b = 1.49$):
users who make more adversarial replies to candidates are more active in general.
It retains its strength and significance across the other two models. In \emph{Model~3}, for example, each additional
reply is related to an increase of $0.13$ in
the number of adversarial replies 
(a standardized coefficient of $b=1.307$, divided by twice the standard deviation
of $5.21$). 
Even after controlling for the overall number of
replies,
the scale of adversarial activity of the user is correlated with other attributes.
For example, a higher number of followers is associated with less adversarial activity. According to \emph{Model~3}, every
tenfold increase in followers is associated with a $0.08$ decrease in 
adversarial activity.
On the other hand, 
there is inconsistent significance for account age and verified status across
the three models, suggesting these are not particularly powerful predictors. 

\begin{figure}[!t]
    \includegraphics[width=\linewidth]{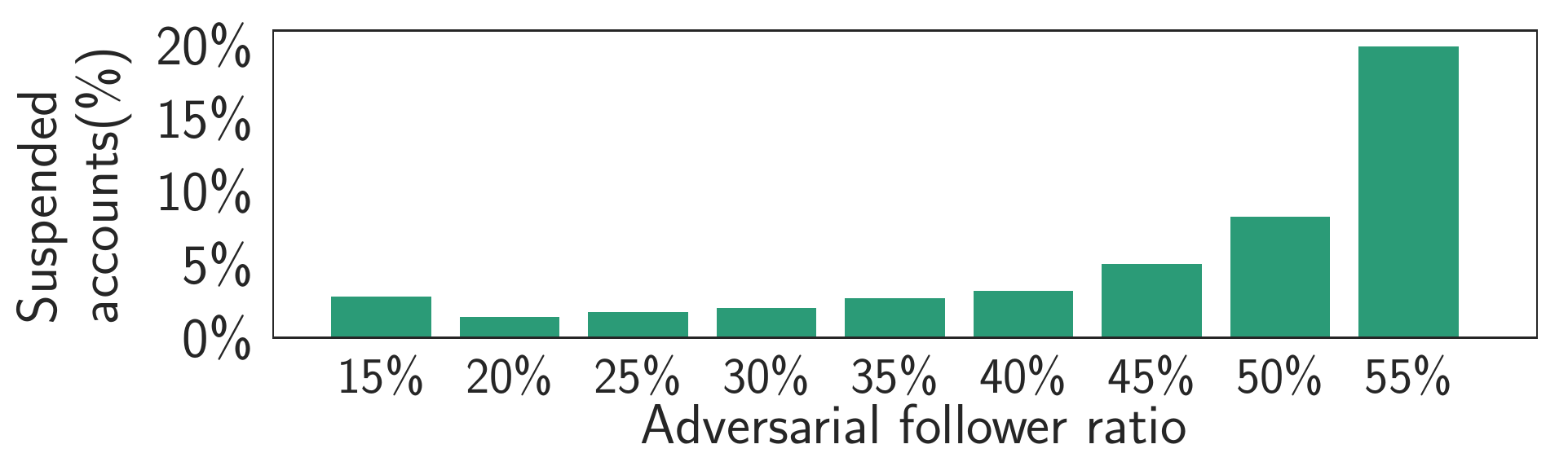}
    \caption{\label{fig:follow} Users with higher adversarial follower ratio are
    generally more likely to have been suspended.} 
    \vspace{-.4cm}
\end{figure}

A user's political activity attributes are strongly correlated with level of adversarial activity. 
The most significant factor is the tendency to engage with candidates from the opposing party on Twitter.
This trend was measured by the number of opponent candidates the user replied to and the percent of
their tweets dedicated to opponent candidates,
both of which correlated with an increase in adversarial activity.
An increase of $1\%$ in the number of replies towards opponent
candidates is correlated with a $0.006$ increase in the quantity of adversarial replies, while 
each additional opponent candidate replied to raises the adversarial activity by~$0.11$. 
Including a partisan hashtag into one's user
profile correlates with a $0.051$ increase in the predicted amount of adversarial activity. 
In contrast, the centrality in the political user network measured by page 
rank is negatively correlated with adversarial activity. 
The less ``important'' users in that network tend to engage in more adversarial interactions.
Other indicators of political activity --- number of retweets and
candidates followed, both associated with partisan support --- 
are also \textit{negatively} correlated with adversarial activity, though with less significance.
For example, according to
\emph{Model~3}, tenfold increases in the number of retweets and each additional
candidate followed are correlated with $0.033$ and $0.001$ decrease in number
of adversarial tweets, respectively.

Finally, \emph{Model~3} shows that adversarial behaviors by friends has strong correlation 
with a user's behavior in posting adversarial posts.  A $1\%$ increase in
friends who posted at least three adversarial tweets is correlated with $0.08$ more
adversarial tweets posted by the user.

Overall, our findings demonstrate a number of key trends correlated with higher
adversarial activity towards candidates on Twitter.
Users who engaged in more adversarial interactions with candidates tend
to have less followers and are less central in the network of users interacting with candidates.
In other words, adversarial activity was higher among ``fringe'' users.
Further, the findings indicate that adversarial activity is correlated with more
attention to opponent candidates and reduced attention to the user's own candidates.
Moreover, adversarial activity is negatively correlated with (likely)
support-oriented activity on Twitter: the more a user retweets or follows
candidates (presumably, of their partisan alignment), the less adversarial
activity they participate in.
Conversely, 
users who pay relatively more attention to and interact
with more candidates from the opposing party produce more adversarial replies. 

\subsection{Accounts followed by Adversarial Users}

The analysis above has shown that 
a user's
adversarial replies to political candidates are associated with both the number
of friends who adversarially interact with candidates and the magnitude of their
friends' adversarial activities. 
In this section we explore whether the behaviors of the accounts followed by disproportionally more adversarial users are indeed more problematic.
We performed an analysis by collecting data about accounts followed by our
\harassers.
Note that these accounts may not necessarily be included in our dataset, as they may have not replied to any House candidate during the data collection period.
In the following comparison, we selected two groups of users from
our dataset: (1) the \textit{Adversarial} group: users who posted at least $3$
adversarial tweets, and  (2) the \textit{Baseline} group: users who posted at least $3$
tweets and never posted adversarial tweets. We define the \textit{Adversarial
Follower Ratio}~(AFR) of an account as the proportion of followers from the
\textit{Adversarial} group over the number of followers from both groups.  In March
2019, we queried Twitter for the user statuses --- i.e. still active on Twitter or not --- of a set of 21 million accounts who are
followed by users in our dataset.  By the time of data collection, $2\%$ of the
accounts had been either deleted or suspended from Twitter. We focus on the
$97$\,K accounts that had at least $100$ followers in total.

Figure~\ref{fig:follow} depicts the rate at which accounts with different
AFR were suspended by Twitter.
Each bin represents at least $100$ accounts.
Interestingly, the chart shows that accounts with higher AFR are more likely to be suspended.
For example, $20\%$ of the accounts with AFR 
in the range \mbox{$50$--$55\%$} were suspended (rightmost bar), which is
significantly higher than the $2\%$ rate of suspension for all collected accounts.
Account suspension on Twitter may result from 
activities like spamming, showing signs of account being compromised, and abusive tweets or behavior~\cite{suspend}.
We have no information about why these accounts were suspended.
It is possible that Twitter is using AFR or similar metrics to prioritize moderation. More likely, the other behaviors of these users that led to suspension were correlated with their AFR.
\section{Analysis of highly adversarial users}\label{sec:qualitative}

\begin{figure}[!t]
    \includegraphics[width=\linewidth]{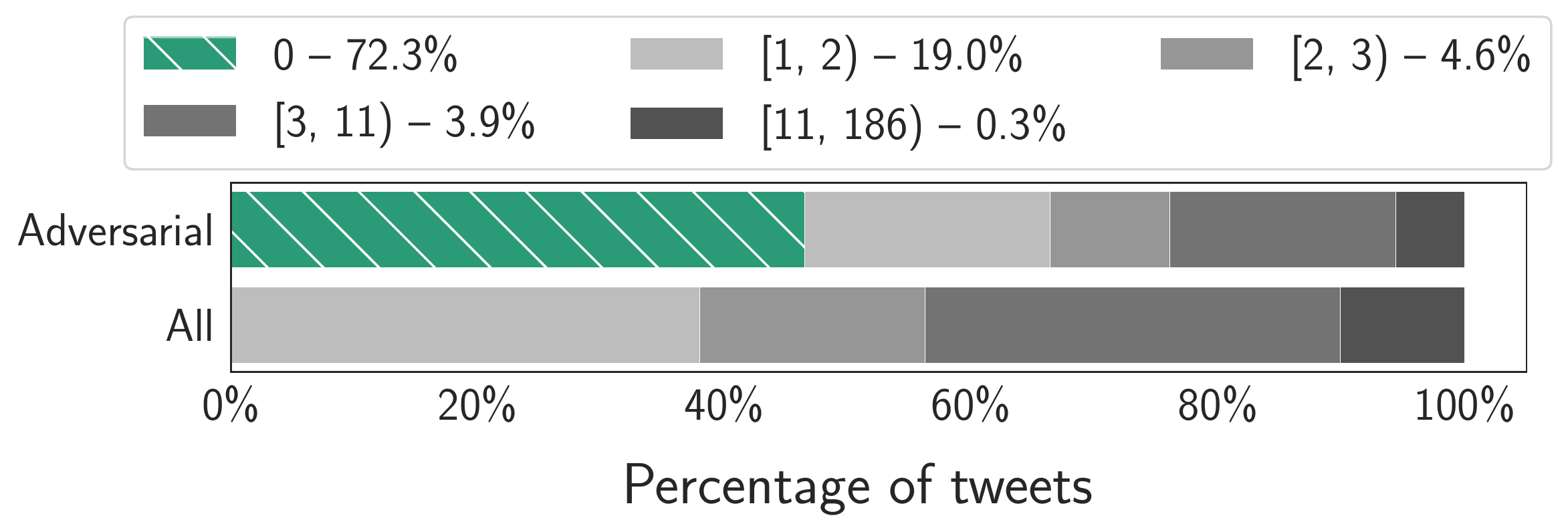}
    \caption{\label{fig:toxicP}Breakdown of tweets by different user groups, as divided by the amount of adversarial
    interactions made~(specified in the legend with group size in percentage).}
    \vspace{-.4cm}
\end{figure}

It is important to consider and understand in more depth the group of highly
adversarial users, defined here as those who made at least $11$ posts that are
$0.7$-adversarial.  Figure~\ref{fig:toxicP} depicts the proportion of all tweets~(top bar)
and of adversarial tweets~(bottom bar) contributed by each user group, where the groups are
based on the number of \textit{adversarial} replies made per user (unlike
Figure~\ref{fig:fig1}b that
grouped users by total reply volume). The highly-adversarial group (dark,
rightmost area of both bars) makes up only $0.3\%$ ($1$,$010$ users) of the
total population of users interacting with the candidates, but these users
contributed $10\%$ of the adversarial replies to candidates (bottom bar). This
group was very active in general, contributing $5.6\%$ of the total replies to
candidates (top bar).

The small size of this user group allows us to use qualitative coding methods to
better understand these users' activities.
We first evaluate the
type of content they produce, in particular in comparison to a control of
replies from other highly-active users.
We also
examine whether the $0.7$-adversarial threshold is in some
way biased in detecting and interpreting the activities of this adversarial
group compared to others.
In addition we perform an analysis based on qualitative coding of the information in the users' profiles,
extending the ``use of political hashtag'' findings from the last section.

\textbf{Content analysis of users' replies to candidates.}
To better understand the activities of highly adversarial users who are responsible for an outsized amount of 
hostile content, we performed qualitative coding of a sample of tweets from both \textit{highly adversarial} and \textit{highly active} users.
The analysis was geared to address two questions:
(1) Does our use of Perspective API bias
our view of who is an adversarial user? 
For example, users may exploit 
different vocabularies or engage in actions that are easier, or harder, for general language models to detect.
(2) Does the content of negative interactions with candidates differ between highly adversarial and highly active users? Moreover, does the content of positive interactions differ?
Note that we use the terminology of negative/positive here instead of adversarial/non-adversarial, as we are not only focusing on hurtful language, 
but rather any kind of negative interaction, including civil disagreement and criticism.
We make the distinction between negative and adversarial interactions as many of the interactions such as attacks against the candidate's policy or political view may not be adversarial,
i.e., not intending to hurt, embarrass or humiliate.
Rather, negative but civil interactions are necessary for democratic discussions.

We engaged in qualitative coding of a sample of 800 tweets from
both user groups.  Specifically, we sampled 400 tweets each from highly
adversarial users with at least 11 $0.7$-adversarial replies in our dataset ($U_{adv}$), 
and highly active users with at least 11 replies in our dataset ($U_{active}$, $19$,$797$ users in total). Note that $U_{adv} \subset U_{active}$.
For each group, we sample 200 tweets that are $0.7$-adversarial and 200 tweets
with adversarial score ${\le}0.7$ (i.e., non-$0.7$-adversarial or ``clean'' according to the threshold-based analysis). We denote these sets of tweets $T_{adv,adv}$ (adversarial users' adversarial tweets) and, accordingly, $T_{adv,clean}$, $T_{active,adv}$ and $T_{active,clean}$.

The coding scheme was developed by the research team in a sequence of coding sessions, where two of the authors independently coded a subset of tweets, then discussed their codes to simplify and streamline the task. We ended up with a coding scheme that asks coders to rate: 
\begin{itemize}[noitemsep,topsep=0pt]
\item \textit{Does the tweet contains civil, mildly uncivil, or severely uncivil content?}
As rater confusion was concentrated around the ``mildly uncivil'' and ``severely uncivil'' categories, we merged them into one ``uncivil'' label in the analysis.
\item \textit{What type of negative interaction is in the tweet, if any?}
Raters are allowed to assign multiple labels including personal attacks, accusations, other attacks against the candidate,
attacks against the candidate's affiliation or leadership,
and attacks against the candidate's policy or political view.
\item \textit{Does the tweet express support for the candidate, the candidate's
party affiliation, or his/her political views/policies?}
Raters are allowed to assign multiple labels.
\item \textit{Is the reply relevant to the original tweet?}
We asked raters to choose a relevance label for each tweet among the following options: irrelevant, same/similar topic, somewhat relevant, and directly responding.
In our analyses we converted the last three labels into a  single ``relevant'' category.
\end{itemize}

Tweets were coded individually using a web-based interface.
In each screen we present coders not only 
the reply tweet to be annotated, but also the original tweet by the candidate for context. In addition, the display included the candidate's political affiliation (Democrat or Republican). 
We did not expand URLs or images in the original or reply tweets. The 800 tweets were assigned to coders in random order.
In total, five different members of the research team labelled $800$ tweets, with each tweet being coded by two team members for a total of $1600$ labels.
We note that, in our estimate, coders exhibit liberal leaning in terms of political preference and have higher than average exposure to U.S. politics, though most of them were U.S. residents but not U.S. citizens.
Familiarity with politics was required to allow the coders to label more complicated
questions, such as whether a tweet contains a political topic,
and whether comments on specific topics might be considered subtly adversarial~\cite{adversarial}.
We excluded all tweets that coders were not able to evaluate, most frequently when a tweet only contained a URL (which we did not expand).
Our final data used in this analysis includes $690$ tweets in total,
$350$ of them come from the highly active group ($T_{active,adv} \bigcup T_{active,clean}$).

\begin{figure}[t!]
    \includegraphics[width=\linewidth]{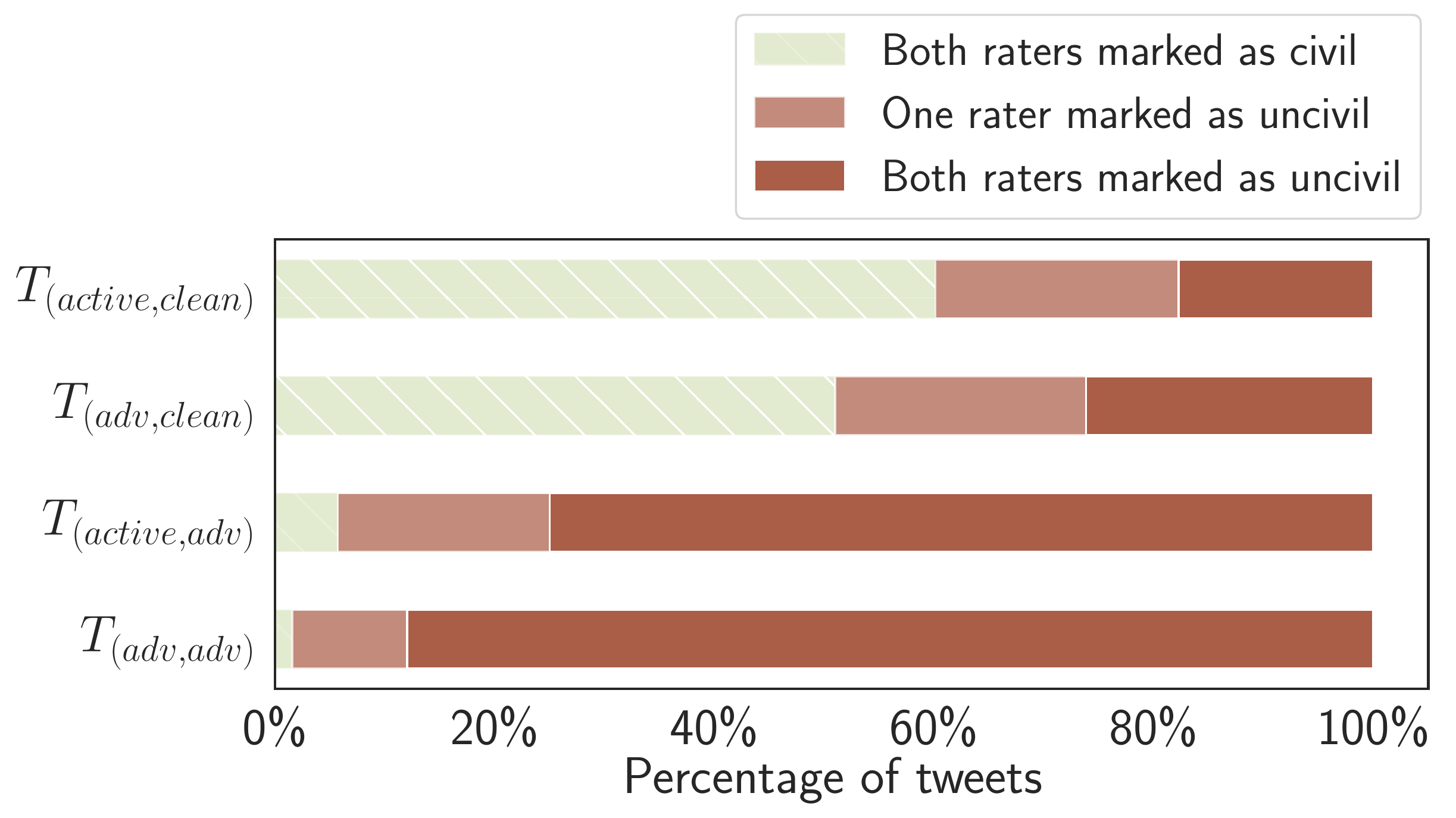}
    \caption{\label{fig:civilResult} Aggregating the scores of two human raters shows significant agreement with the automatic classification results.}
    \vspace{-.5cm}
\end{figure}
\begin{figure*}[!t]
  \includegraphics[width=\textwidth]{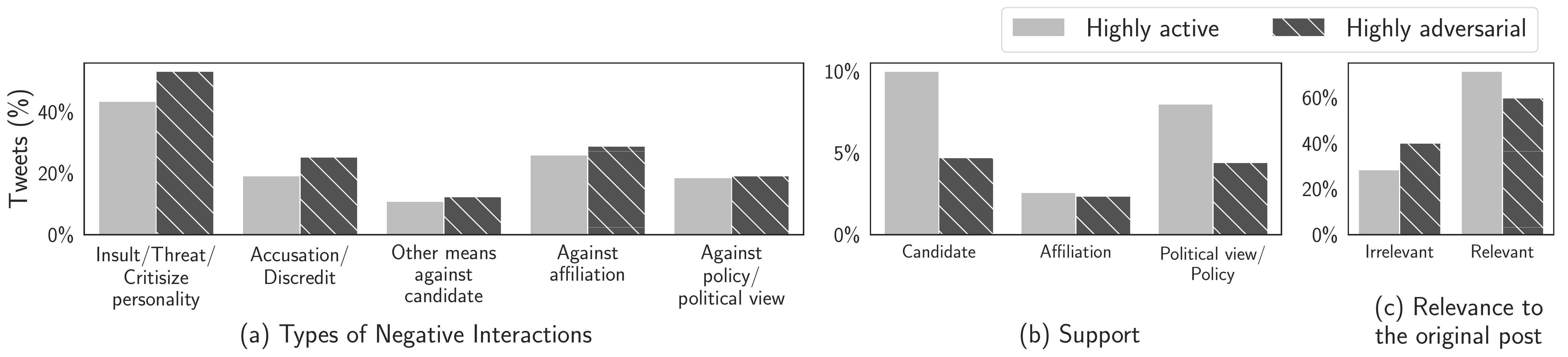}
\caption{\label{fig:codedres}Highly adversarial users exhibit different activity patterns in terms of \textbf{(a)} types of negative interactions and \textbf{(b)} types of supportive interactions they engaged in,
as well as \textbf{(c)} reply relevance to original posts compared to highly active users.}
\end{figure*}

Human annotation exhibited significant agreement with algorithm-labeled adversarial-ness.
The raters were reasonably consistent in evaluating whether a given reply to a candidate contained content that can be labeled as ``civil'' or ``uncivil'', with $0.6$ based on Cohen's Kappa, high end of the moderate agreement range.
Figure~\ref{fig:civilResult} aggregates the \textit{civility} ratings.
Tweets that were given ``uncivil'' scores by both raters are marked in dark red while those that were given ``civil'' scores by both raters are marked in light green and with a stripe pattern.
Tweets that were assigned with disagreeing labels are marked in light red.
The rows represent the sets of tweets analyzed based on the group that produced them (active or adversarial users), and whether the tweet was marked as adversarial by the algorithm. From the top, the rows represent $T_{active,clean}$, $T_{adv,clean}, T_{active,adv}$, and $T_{active,adv}$. 

The figure indicates the differences between the two groups, as well as the performance of the $0.7$-adversarial threshold method. 
First, the chart indicates that $0.7$-adversarial tweets from both groups (bottom two rows) are indeed often labeled as adversarial, with over 75\% of the tweets in either row marked as ``uncivil'' by both raters, and close to 100\% marked as ``uncivil'' by at least one rater. 
The trend is opposite for the top two rows, showing that, while some tweets may be missed by the algorithm, it is not likely to affect the classification of adversarial users --- the algorithm may miss more of their tweets compared to other users' adversarial tweets.
Taken together, we conclude that the $0.7$-adversarial threshold method is an effective means to identify highly adversarial users.
In the following discussion,
we take a deeper dive into a detailed analysis on the diversity of tweets posted by these users.

We now compare the type of contents in our sample of tweets by adversarial users ($T_{adv,adv} \bigcup T_{adv,clean}$) and our sample of tweets by active users ($T_{active,adv} \bigcup T_{active,clean}$) in Figure~\ref{fig:codedres}.
For each tweet, we take the union of labels assigned by raters.
We use $\chi^2$ tests to compare the two groups, running tests separately for each category as the categories are not mutually exclusive; the $p$-values we report on are significant even after correcting for multiple tests. 
Note that similar to the above analysis in Figure \ref{fig:civilResult},
our results likely underestimate the magnitude of the difference between the groups, as the highly active sample is drawn from a set that also contains the highly-adversarial users as well as other users who are adversarial but did not make the 11-tweet threshold for~$U_{adv}$.
Moreover, in this analysis we use a \textit{balanced sample} of tweets of each kind (clean or adversarial) made by each group of users. In other words, we are not just comparing the groups' overall activities, we are directly comparing ``clean'' and ``adversarial'' tweets made by each group, showing the differences in types even within these categories.

Figure~\ref{fig:codedres}a shows the manually-labeled categories
assigned to the users' tweets. 
It is clear that the users in $U_{adv}$ are more
likely to engage in personal attacks against candidates (leftmost bar, $\chi^2(1,N=690)=6.25$, $p < 0.05$). 
On the other hand, Figure~\ref{fig:codedres}b shows that $U_{active}$ users are more likely to engage in supportive interactions with candidates (of their own partisan affiliation, naturally).
Particularly they express support more frequently to the candidate themselves~($\chi^2(1, N=690)=6.31$, $p < 0.05$). 
Finally, Figure~\ref{fig:codedres}c shows 
that when considering all tweets with agreeing labels, highly active users were more likely to respond in a manner that is relevant to the candidate tweet that they are replying to~($\chi^2(1,N=492)=6.95$, $p < 0.01$).

We also compared the sets of candidates that $U_{adv}$ and $U_{active}$ replied to,
as well as the sets these groups posted adversarial replies to. Both groups seem to focus on the most
``visible'' candidates. There were no significant differences in
the set of candidates being replied to between the two groups.

In summary, the results presented in Figure~\ref{fig:codedres} expose different types of adversarial engagements and interactions performed by the highly-adversarial users. 
In particular, adversarial users are more likely to engage in personal attacks and are \textit{less} likely to engage in other supportive interactions or respond with relevant content with candidate tweets.
In other words, the difference in motivations of the adversarial users is manifested in all their behaviors.

\textbf{Analysis of user profiles.}
Table~\ref{tab:regression} shows indications that the profiles of
highly adversarial users' accounts may differ from that of other groups ---  specifically, the presence of a partisan hashtag in the user profile has significant positive association with the amount of adversarial interaction with political candidates.
To understand and elaborate on this trend we performed a qualitative coding study of Twitter user profiles, to examine how partisanship is represented differently in adversarial users' profiles. In addition, we examined whether these profiles are more likely to appear anonymous. 

In particular, we randomly selected 100 user profiles from each of the following
three groups: (1) \textit{random users}: a set of users chosen at random (denoted~$U_r$);
(2) \textit{highly adversarial}: users who have posted at least 11 adversarial
tweets ($U_{adv}$ as defined above); and
(3) \textit{active non-adversarial}: 
a subset from \textit{highly active} users, who have posted at least 11 tweets and never posted any adversarial tweet ($U_{active,clean}$, $4$,$867$ users in total). 
The last group is different than $U_{active}$ in the previous discussion from Figure~\ref{fig:civilResult} in that it specifically excludes users posting adversarial content.
We developed a coding scheme in a sequence of coding sessions, where two of the authors coded a subset of user profiles, then discussed their codes to focus on the key differences between accounts in respect to this task. Our coding scheme asks coders to rate user profiles as: 

\begin{figure}
  \includegraphics[width=\linewidth]{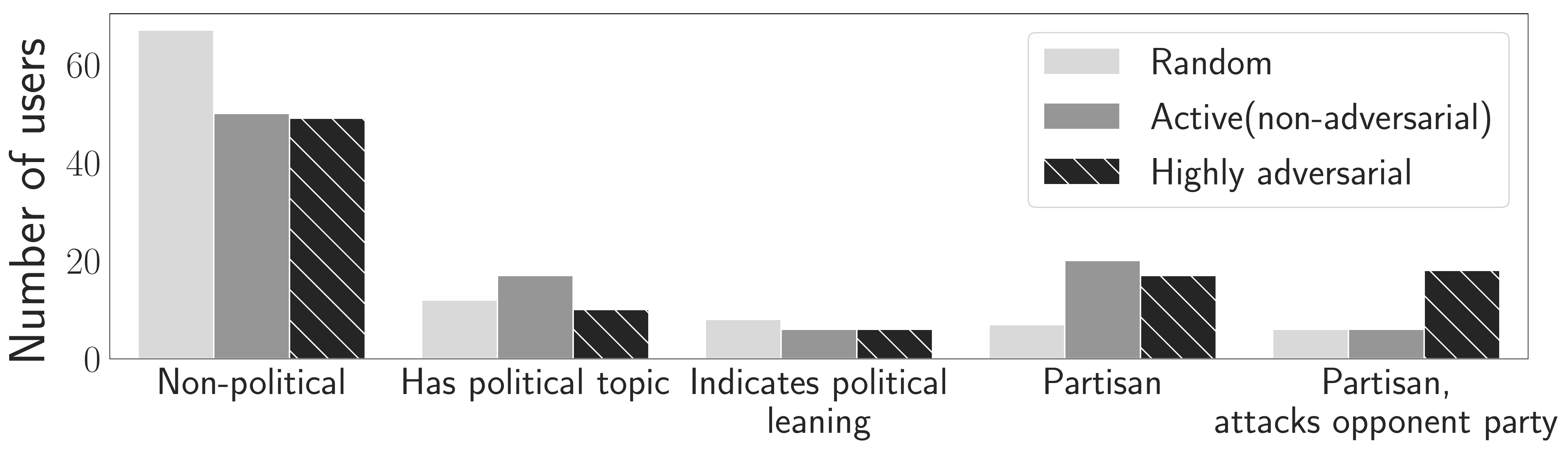}
  \caption{\label{fig:description} Highly adversarial users are more likely to
  include attacks on the opposite party in their user profiles.}
\vspace{-.5cm}
\end{figure}

\begin{itemize}[noitemsep,topsep=0pt]
\item \textit{Non-political}: A profile without political topics or empty.
\item \textit{Mentions political topic}: User profiles that mention interest in political topics, but do not suggest political leaning. 
\item \textit{Shows political leaning}: User profiles that mention political views and opinions~(``liberal'', ``Blue Lives Matter'') that do not explicitly include partisan support.
\item \textit{Partisan}: User profiles explicitly expressing partisan support~(``\#MAGA'', ``\#BlueWave'').
\item \textit{Partisan attack of opponents}: User profiles attacking the political opponent, their party or leadership~(``Anti Trump'').
\end{itemize}

Two annotators independently coded each profile of our sample of 300 accounts. The rating was based on the user profile details including Twitter handle, profile description, location and name. 
Disagreements were resolved by a third annotator.
One of the user profiles was excluded from the data as no consensus was reached.

Figure~\ref{fig:description} shows the distribution of each category of user
profiles for different groups.
The histogram shows the categories on the X-axis, and the number of users of each group whose profile is in each category (Y-axis). 
The three groups are significantly different~($\chi^2(8,N=299)=21.99$, $p < 0.005$). 
For example, the random users were more likely (lightest column in the leftmost group) to have non-political accounts, though a majority of accounts across all the three groups were in that category. 
On the other hand, the active and
adversarial users were more likely to show partisan support.
However these differences are not significant in a post-hoc analysis.
More significantly, highly adversarial users were more likely to express their
opposition to the opposing party in their user profile (the rightmost dark column, $\chi^2(2,N=299)=10.56$, $p < 0.01$).

We also explored whether highly-adversarial users tend to use accounts that appear anonymous~(i.e. without clear identifying information) more
frequently than the other two groups. To this end, we coded every profile as
``presenting as anonymous'' or not. In other words, based on the profile handle,
description, and published name, we coded as ``anonymous'' those profiles who
cannot be connected to a specific individual (note that we naturally cannot
detect users who offer identifying details that are not correct). While 48 out
of the 100 random users were labeled as anonymous, the active and the
high-adversarial groups had 63 out of 100 and 64 out of 100 anonymous profiles, respectively. 
The difference between these groups was statistically significant ($\chi^2(2,N=300)= 6.6103$, $p < 0.05$).

\section{Discussion}

Our results provide some reflection on earlier findings on adversarial behaviors in social media.
One hypothesis from the literature is that arbitrary users can turn adversarial
given an appropriately triggering context~\cite{cheng2017anyone}. Our results do
not refute (or support) this hypothesis, but we do see an alternative pattern of
behavior: users that seem to seek out opportunities for adversarial
interactions. Two aspects of our analysis can be interpreted as supporting
this viewpoint. First, highly adversarial users reply more in
general to candidates from the party the user opposes, suggesting that they are
actively seeking out these opportunities for cross-party communications. Second,
highly adversarial users are significantly more likely to have profiles that
include partisan attacks, suggesting an adversarial disposition independent of
any specific conversation.

Future research on improving online conversations can build upon our observations.
For example, our findings from both qualitative and quantitative analyses show that users with
higher adversarial activity level engage in fewer supportive interactions with
political candidates.
One possible direction as a future study is to further understand if
encouraging supportive interactions would decrease user participation in 
adversarial behaviors.
In addition, 
our analysis shows that users with lower activity level have a higher participation in 
adversarial behaviors as compared to their general posting activities.
While prior research often focus on how to encourage user engagement from inactive users~\cite{nonnecke2001lurkers,tedjamulia2005motivating},
it is as important to understand how to encourage \textit{high-quality} engagement.

Our findings offer design implications for tools assisting human-in-the-loop content moderation in social media.
For example, our results show that highly adversarial users are more likely to attack the opponent party in their user profiles,
suggesting that tools can be developed to surface users who have expressed negativity explicitly in their profiles.
This would enable moderators to prioritize reviewing content generated by these users.
Further, social media platforms could take more granular measures to discourage users from seeking out conflicts.
For example, down-ranking toxic interactions with public figures can both limit public exposure to the adversarial behaviors and create more friction for conflict-seeking users.

Finally, we urge platforms to take stronger actions to help political figures against adversarial interactions on social media,
because of the ramifications for our democracy and society.
In line with prior findings on social media participation~\cite{preece2004top,nonnecke2000lurker},
``repeat'' offenders contributed a large proportion of the adversarial contents.
These users might be good targets for punitive actions by the platform, such as banning.
Yet in our measurements we saw that only $15\%$ of the highly adversarial users were later banned by Twitter.

\textbf{Limitations.} Our measurement study naturally
suffers from some limitations. 
First, our study, being observational, cannot speak to the causal relationships between the variable we studied.  
Second, our focus was on the U.S.~political
system, and in particular the elections for the U.S.~House of Representatives;
the results may not generalize to other settings. Likewise, our findings may
also be skewed by the specific topics which were in the news during the data
collection period. For example, the confirmation hearings for U.S.~Supreme Court
justice Brett Kavanaugh were at the center of attention in that time period, and may
have triggered more or less adversarial interactions than would be observed at
other periods of time. 
In fact, as we pointed out, previous studies on characterizing adversarial behaviors 
under various scenarios on Twitter~\cite{ribeiro2018characterizing,chatzakou2017mean,chatzakou2017measuring,elsherief2018peer} reported differing trends on certain features in term of correlations with adversarial activities.
For example, ~\cite{chatzakou2017measuring} shows that the number of followers a user has is positively correlated
with adversarial behaviors while ~\cite{chatzakou2017mean,ribeiro2018characterizing} and our study suggest the opposite trend.
This observation might suggest that adversarial behaviors are contributed by different social media populations and models need to be developed tailered to the context.
Therefore these results are not ready to be applied directly on fully automated detection.

Third, our qualitative analysis is based on answers from
raters who exhibit liberal leaning.
Recent research~\cite{joseph2019polarized} has shown that
although people with different political leanings disagree on the severity of uncivil tweets,
they agree when distinguishing uncivil tweets from civil ones.
Nevertheless, it is possible that a group of coders with more diverse political backgrounds could bring a different perspective to the analysis.
Fourth, when annotating tweets, coders were exposed to the political affiliation of the politicians.
Although this design was intentional in order to improve annotation quality, as many tweets are hard to evaluate without this context, we acknowledge that this may introduce bias into our analysis.

Finally, as previous works have pointed
out~\cite{davidson2017automated,adversarial}, harassment detection
based on machine learning tends to focus on usage of offensive language, while having a high false
negative rate when classifying subtle adversarial content.
Further, the tool used in our analysis, Perspective API, does not include Twitter in its training data.
While we validated
the precision of our method in this and earlier work~\cite{adversarial}, 
our method may nevertheless misses or ignores uncivil content not being flagged
and underestimate the amount of adversarial activities. 
In fact, we showed in our qualitative analysis that our approach is more likely to \textit{underestimate} the amount of adversarial activities produced by highly adversarial users, which has minimal impact for our results.
\section{Conclusions}

We studied users who engage in adversarial activities against political candidates,
with a dataset collected during the run-up to the 2018 U.S. midterm elections.
Using existing machine learning tools,
we found that while a significant
number of adversarial replies are made by those who
tweeted to candidates very few times over the data collection period, 
a small minority of active users is responsible for an outsized portion of adversarial
interactions.  
Using a combined approach with both quantitative and qualitative analysis that used features specific to the context of U.S. politics, 
we showed that for moderately active users, those who are more adversarial are both less
central in the social graph consisting of users who interact with candidates, 
and pay less attention to candidates of their own partisan affiliation.
When compared to highly active users, 
highly adversarial users tend to focus their
activities more on negative interactions with candidates and less on supportive
ones.
Our results also indicate that 
highly-adversarial users are more likely to express 
negativity towards political opponents in their user profiles.
With these findings we hope to inform future research in this area
and encourage platform efforts in protecting 
political figures from adversarial interactions.
\section{Acknowledgement}

We thank Jingxuan Sun, Xiran Sun and Dinesh Malla for their help annotating the data.
We thank Andreas Veit and Maurice Jakesch for their feedback and support.
This research is supported by NSF research grants CNS-1704527 and IIS-1665169, as well as a Cornell Tech Digital Life Initiative Doctoral Fellowship.

\balance{}

\bibliographystyle{SIGCHI-Reference-Format}
\bibliography{chi2020}

\end{document}